%% file: main.tex
\documentclass[a4paper,11pt]{article}
\usepackage{pos}

\usepackage{xspace}
\usepackage{hyperref}
\usepackage{wrapfig}
\usepackage{enumitem}
\usepackage{lineno}

\newcommand{\Xmax}{\ensuremath{X_\text{max}}\xspace}
\newcommand{\sib}[1]{\textsc{Sibyll}\,#1\xspace}
\newcommand{\qgs}{\textsc{QGSJet}\xspace}
\newcommand{\qgsii}{\textsc{QGSJet~II-04}\xspace}
\newcommand{\qgsiii}{\textsc{QGSJet~III-01}\xspace}
\newcommand{\epos}{\textsc{Epos}\xspace}
\newcommand{\eposlhc}{\textsc{Epos-LHC}\xspace}
\newcommand{\eposlhcr}{\textsc{Epos-LHC-R}\xspace}
\newcommand{\gcm}{\,\ensuremath{\text{g}/\text{cm}^{2}}}
\newcommand{\Rhad}{\ensuremath{R_\text{had}}\xspace}
\newcommand{\RhadTheta}{\ensuremath{\Rhad(\theta)}\xspace}
\newcommand{\DeltaXmax}{\ensuremath{\Delta\Xmax}\xspace}
\newcommand{\avg}[1]{\ensuremath{\langle{#1}\rangle}}

\title{Update on testing of air-shower modelling using combined data of the Pierre Auger Observatory and phenomenological consequences}
 \ShortTitle{Update on testing of air-shower modelling}

\author*[a]{Jakub Vícha}

\affiliation[a]{FZU - Institute Of Physics of the Czech Academy of Sciences, Prague, Czech Republic}

\onbehalf{for the Pierre Auger Collaboration$^b$}
\affiliation[b]{Observatorio Pierre Auger, Av.\ San Mart{\'\i}n Norte 304, 5613 Malarg\"ue, Argentina\\
Full author list: {\rm\url{https://www.auger.org/archive/authors_icrc_2025.html}}}

\emailAdd{spokespersons@auger.org}

\abstract{The combined data of Fluorescence and Surface Detectors of the Pierre Auger Observatory has recently provided the strongest constraints on the validity of predictions from current models of hadronic interactions. The unmodified predictions of these models on the depth of shower maximum (\Xmax) and the hadronic part of the ground signal are unable to accurately describe the measured data at a level of more than 5$\sigma$ in the energy range 3-10 EeV. This inconsistency has been shown to originate not only from the predicted amount of muons at the ground level, but also from the predicted scale of \Xmax, which must be adjusted to better match the observed data. The resulting deeper \Xmax scales of the models imply a heavier mass composition to be interpreted from the \Xmax measurements.

We show the results of the test with an updated data set of the Pierre Auger Observatory, studying also the energy evolution of the fitted modification parameters and new versions of the models of hadronic interactions. Additionally, we discuss the phenomenological consequences of the deeper \Xmax scale of models on the interpretation of the features of the energy spectrum and the muon problem in air-shower modelling.}

\ConferenceLogo{icrc_logo}

\FullConference{%
39th International Cosmic Ray Conference (ICRC2025)\\
15 -- 24 July, 2025\\
Geneva, Switzerland}

\begin{document}
\maketitle

\section{Introduction}
The combined data from the Surface and Fluorescence Detectors of the Pierre Auger Observatory \cite{PACosmicObservatory} allow to put strong constraints on the predictions of models of hadronic interactions.
The method of fitting two-dimensional histograms of the ground signal at 1000\,m from the shower core and the depth of shower maximum, \Xmax, has shown, for the first time, 5$\sigma$ tension between the prediction of hadronic interaction models \eposlhc \cite{EposLHC}, \qgsii\cite{Qgsjet}, \sib{2.3d}\cite{Sibyll} and measured data for energies 3-10\,EeV \cite{PierreAuger:2024neu}.
The model predictions were assumed to be modified by mass and energy independent parameters \DeltaXmax and \RhadTheta, shifting the predicted \Xmax scale and rescaling the hadronic part of the ground signal at 1000\,m from the shower core, respectively.
In this way, we removed the main differences between the model predictions.
The data were shown to be best described when not only the hadronic part of the ground signal was increased by $\approx$(15-25)\%, but also the predicted \Xmax scale was shifted deeper by $\approx$(20-50)\gcm.
The fitted primary fractions of four primary species: protons (p), helium (He), oxygen (O), and iron (Fe) nuclei combine to a heavier mass composition than is usually estimated from the \Xmax fits to the unmodified model predictions \cite{Auger-LongXmaxPaperMass}.
We have also shown that in the case of the \qgsii model, there is a strong indication of too hard muon spectra generated by the model at 1000\,m from the shower core.

In this proceedings, we present an update of the method applied to newer data than in \cite{PierreAuger:2024neu} with an extended energy range $10^{18.4-19.5}$\,eV,  and we test new versions of the models of hadronic interactions: \eposlhcr \cite{Pierog:2023ahq}, \qgsiii \cite{QGSJetIII} and slightly modified model \sib{2.3e}.
We also test for the energy dependence of the modification parameters and indicate phenomenological consequences about the energy spectra and muon scale.

\section{Testing New Models of Hadronic Interactions}
We follow the high-quality selection of combined Surface and Fluorescence Detector data as in \cite{PierreAuger:2024neu} extended by about 20\% more events in the energy range $10^{18.5-19.0}$\,eV, collected from $1^\text{st}$ January 2004 up to $31^\text{st}$ December 2021.
On top of this benchmark energy range with 2740 events, we extend our analysis to energy ranges $10^{18.4-18.5}$\,eV, $10^{18.5-18.7}$\,eV, $10^{18.7-19.0}$\,eV and $10^{19.0-19.5}$\,eV with 1407, 1670, 1070 and 516 events, respectively, to study a possibility of an energy dependence of modification parameters.
Monte Carlo simulations were produced using CORSIKA 7.8010 \cite{CORSIKA,CORSIKA2,CORSIKA3} and the detector simulation and shower reconstructions were processed using the Auger Offline code \cite{Offline}.

\begin{figure}
    \includegraphics[width=0.5\textwidth]{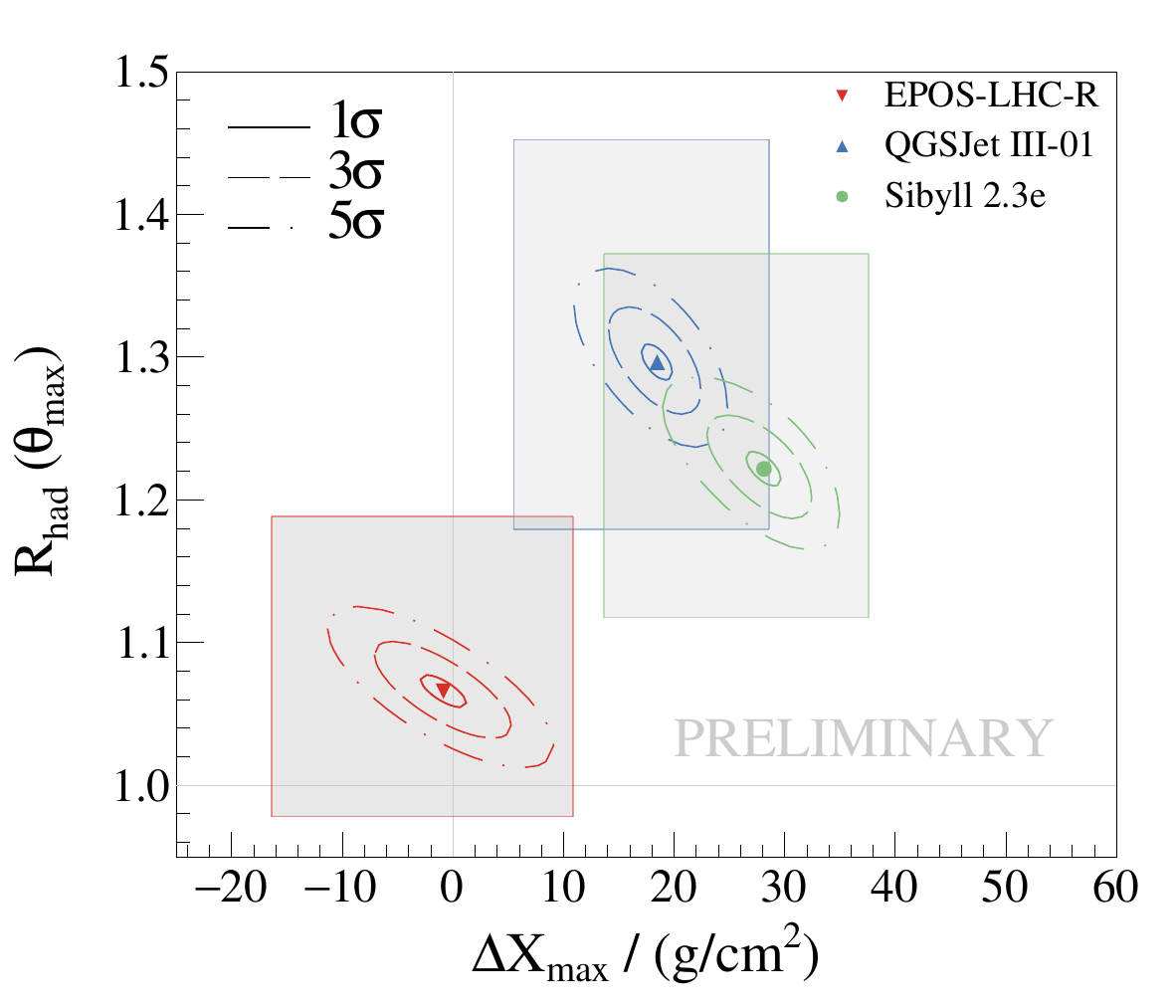}
    \includegraphics[width=0.5\textwidth]{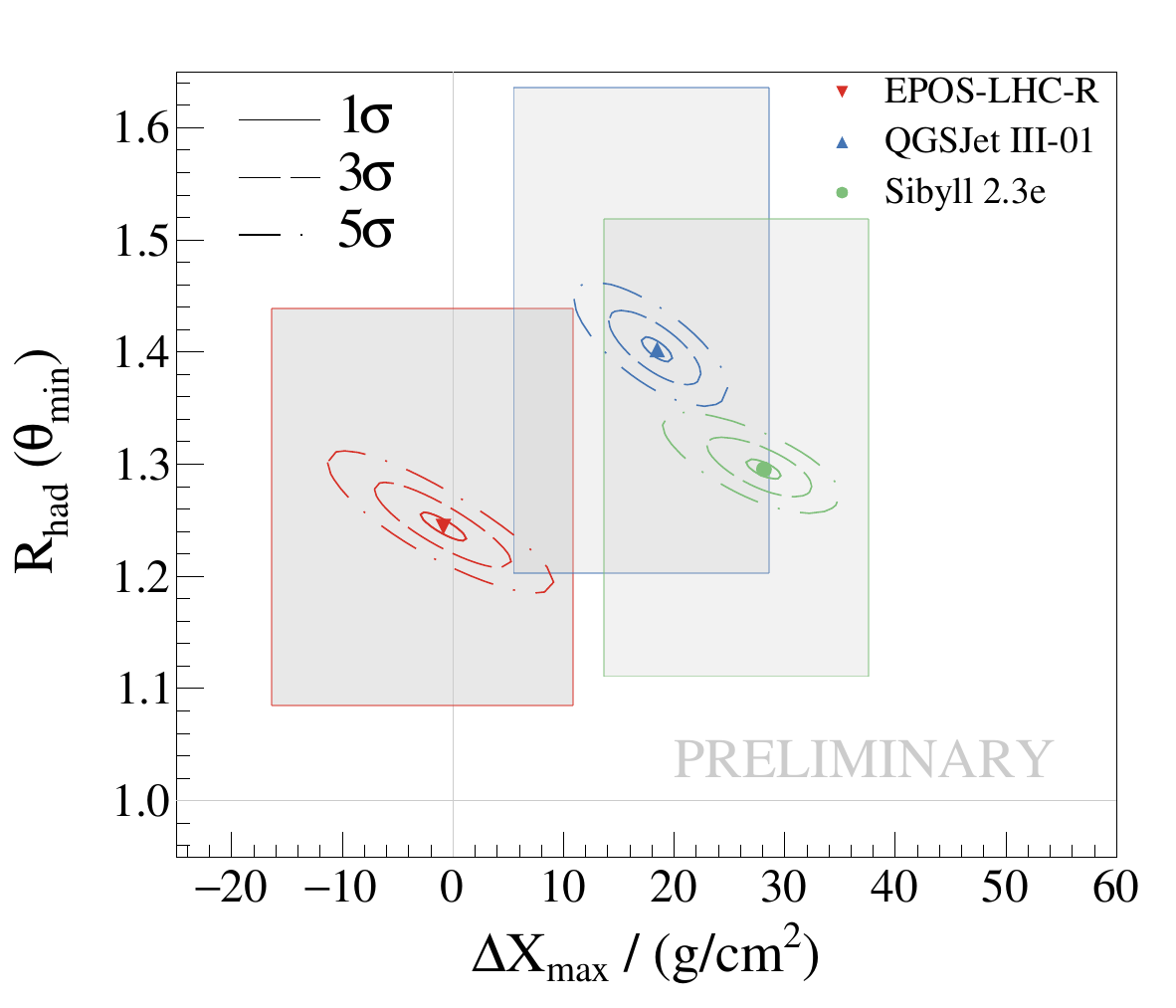}
    \caption{Correlations between \DeltaXmax and $R_\text{had}(\theta_\text{max}\approx55^{\circ}$) (left) and $R_\text{had}(\theta_\text{min}\approx28^{\circ}$) (right) modifications of the model predictions obtained from the data fits in the energy range $10^{18.5-19.0}$\,eV. The contours correspond to 1$\sigma$, 3$\sigma$, and 5$\sigma$ statistical uncertainties. The gray rectangles are the projections of the total systematic uncertainties.}
    \label{fig:RhadVsDeltaXmax}
\end{figure}

\begin{figure}
    \includegraphics[width=0.5\textwidth]{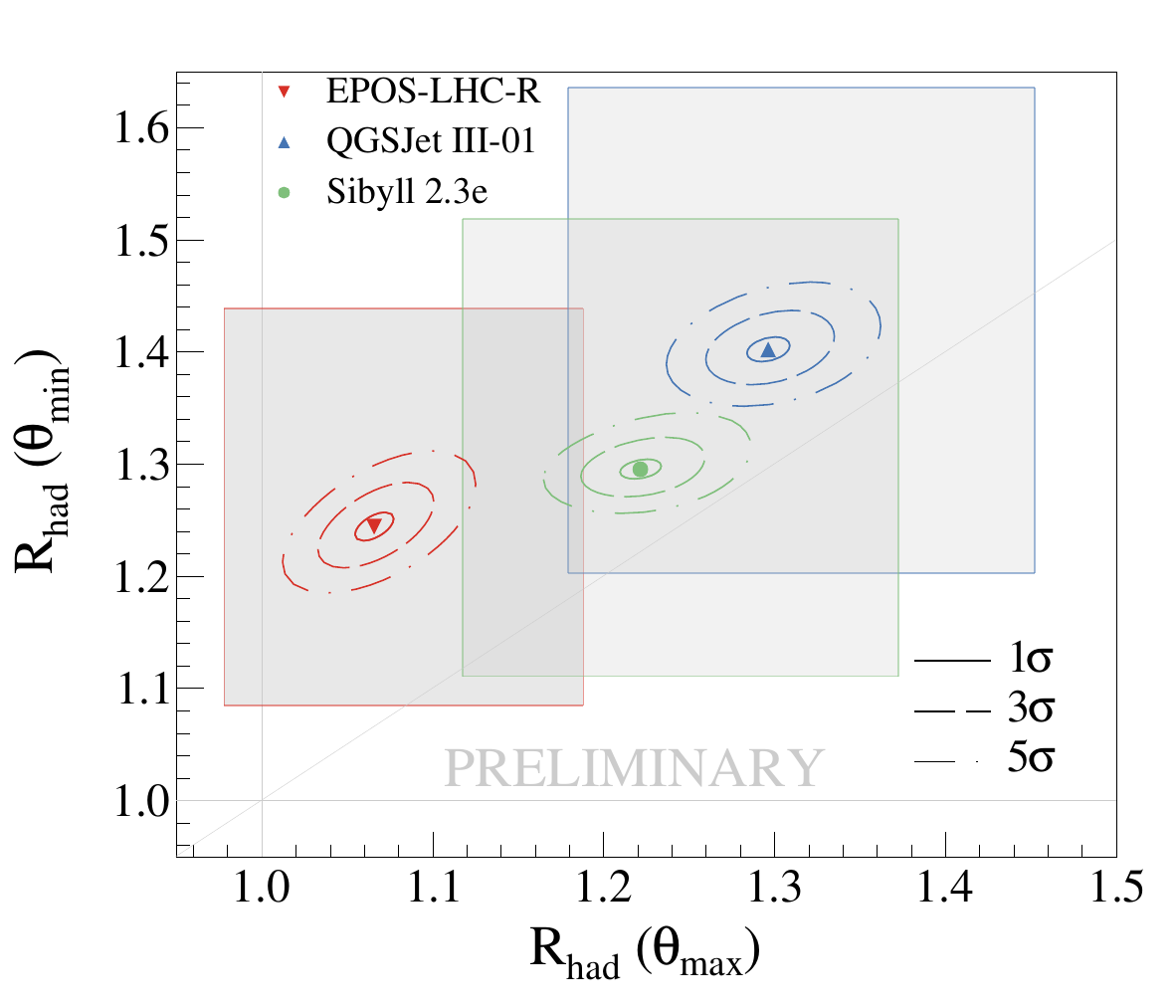}
    \includegraphics[width=0.5\textwidth]{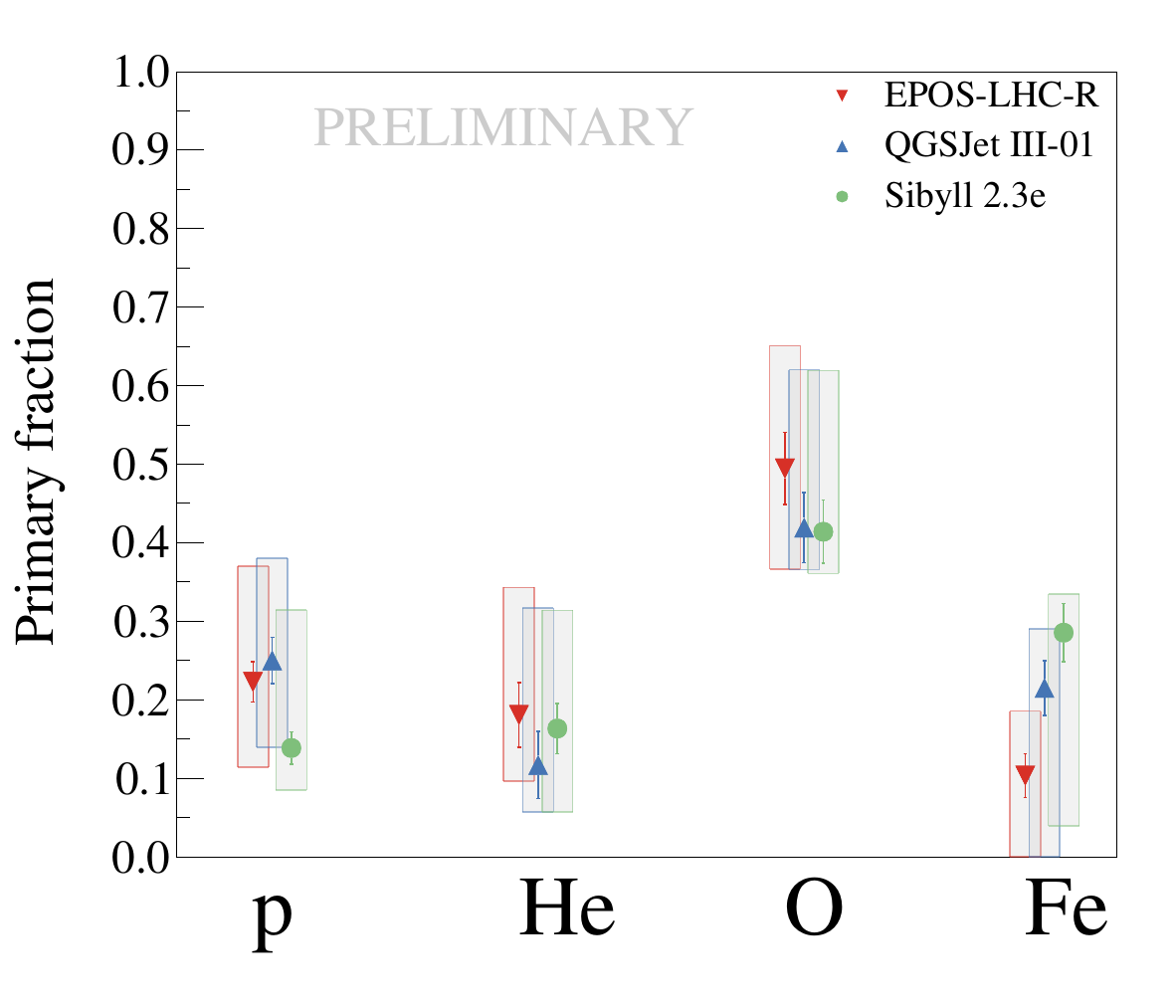}
    \caption{Left panel: Correlations between $R_\text{had}(\theta_\text{max}\approx55^{\circ}$) and $R_\text{had}(\theta_\text{min}\approx28^{\circ}$) modifications of the model predictions obtained from the data fits in the energy range $10^{18.5-19.0}$\,eV. The contours correspond to 1$\sigma$, 3$\sigma$, and 5$\sigma$ statistical uncertainties. The gray rectangles are the projections of the total systematic uncertainties. Right panel: The most likely primary fractions of the four components from the data fits using \DeltaXmax and \RhadTheta. The height of the gray bands shows the size of projected total systematic uncertainties.}
    \label{fig:RhadsAndPrimFractions}
\end{figure}

We show in Fig.~\ref{fig:RhadVsDeltaXmax} and Fig.~\ref{fig:RhadsAndPrimFractions} the resulting modification parameters of the simulated templates, and their correlations, after application of the log-likelihood fit described in \cite{PierreAuger:2024neu} for the benchmark energy range.
The primary fractions (see the right panel of Fig.~\ref{fig:RhadsAndPrimFractions}) obtained using the new model versions are compatible with the values found for the older versions in \cite{PierreAuger:2024neu}, despite the large differences in predictions of the old and new versions of models \qgs and \epos.

The model \sib{2.3e} shows compatible value of \DeltaXmax as for \sib{2.3d} in \cite{PierreAuger:2024neu}, as expected, but the needed rescaling of the hadronic part of the ground signal is now by 5-10\% larger, mainly as a consequence of improvements applied in the reconstruction of the Surface Detector signal.

There are large changes in the predictions on air-shower properties in the case of the updated \qgs model, including deeper \Xmax predictions for protons by $\approx15\gcm$, while for iron nuclei by $\approx25\gcm$.
The \Xmax fluctuations for iron nuclei in this model are now at the level of the total defragmentation of the nucleus, which was previously pointed out by \cite{OstapchenkoXmaxWidth} as a bug in the model \eposlhc, and was fixed in the new \epos version.
As a consequence, the fitted \Xmax shift for \qgsiii is now smaller than in \cite{PierreAuger:2024neu} for \qgsii, being at the level of 20\,\gcm only, however, the hadronic signal at 1000\,m needs to be increased by a larger value at the level of 30-40\%.

The best performing model, in general, is the \eposlhcr model with the predicted \Xmax scale compatible according to the test.
However, there are large differences in the hadronic rescaling at the two extreme zenith angles (see the left panel of Fig.~\ref{fig:RhadsAndPrimFractions}), strongly indicating that the predicted muon spectra at 1000\,m from the shower core are too hard than what is measured, and the hardest among the studied models.

\begin{figure}
    \centering
    \includegraphics[width=0.8\textwidth]{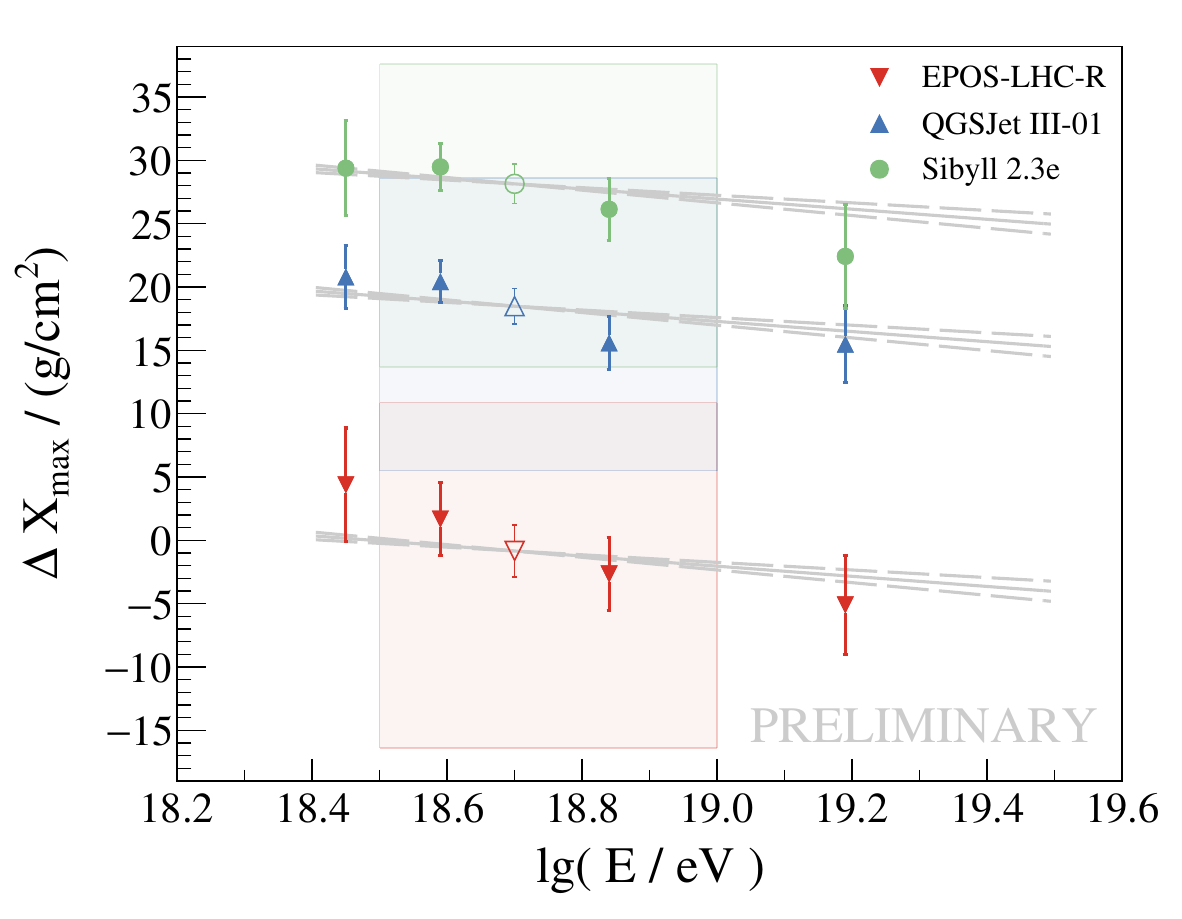}
    \caption{The fitted values of the \DeltaXmax parameter for various selections of energy range. The benchmark energy bin (open markers) contains also the bands of the projected systematic uncertainties. The expected systematic effect of the method $(4\pm1)$\gcm per decade of energy is shown by gray lines.}
    \label{fig:DeltaXmaxVsEnergy}
\end{figure}

We have also tested possible energy dependence of modification parameters, see Fig.~\ref{fig:DeltaXmaxVsEnergy} for \DeltaXmax, by dividing the measured data into multiple ranges of energy.
Given the available event statistics, the benchmark values of \DeltaXmax obtained in the energy range $10^{18.5-19.0}$\,eV were found compatible with the values for other energy ranges for all three models. Similar results were found for the two values of \RhadTheta.
This finding supports the assumption of mass-independent modification parameters, which would be otherwise expected to manifest through the energy-per-nucleon scaling as a consequence of the Superposition model \cite{SemiSuperpositionModel}.
Our specific searches for mass dependencies of modification parameters further support this claim. 
From data on \DeltaXmax we can not also exclude a mild energy dependence in the studied energy range given the available statistics. 
However, such an effect is expected for the energy evolution of \avg{\ln A} by $\approx1$ \cite{Auger-LongXmaxPaper} between $10^{18.4-19.5}$\,eV, which affects the mass-dependent bias on \DeltaXmax coming from the method itself as it was shown in Fig.~12 in \cite{PierreAuger:2024neu} for the older versions of models.
We illustrate estimation of such an effect using gray lines corresponding to the change of \DeltaXmax bias by $4\pm1$\,\gcm per decade of energy in Fig.~\ref{fig:DeltaXmaxVsEnergy}.

\section{Phenomenology Consequences of Deeper Scale of \Xmax}
\begin{figure}
    \includegraphics[width=1.0\textwidth]{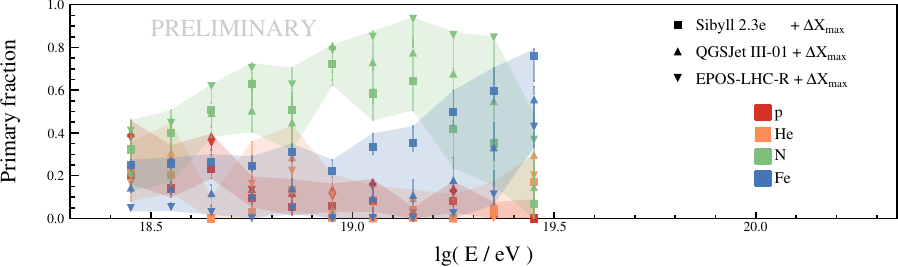}\\[0.5cm]
    \includegraphics[width=1.0\textwidth]{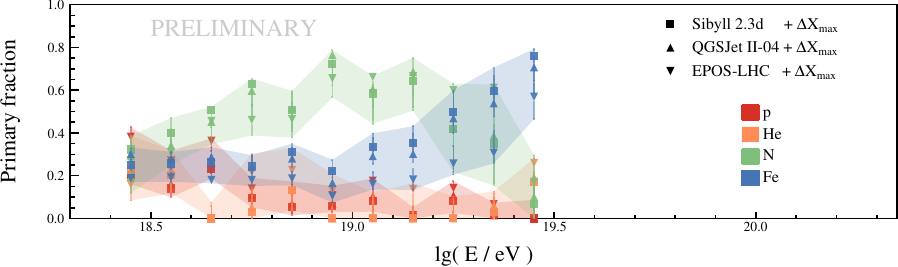}
    \caption{The energy evolutions of four primary fractions fitted to the \Xmax distributions using modified templates by \DeltaXmax for older (bottom panel) and new (top panel) versions of the hadronic interaction models.}
    \label{fig:PrimFractionsVsEnergy}
\end{figure}

Given the non-observation of dependencies of modification parameters on energy or mass, we show in Fig.~\ref{fig:PrimFractionsVsEnergy} the primary fractions obtained by fitting the \Xmax distributions \cite{AugerMassICRC25} using modified simulation templates by \DeltaXmax from the Table~\ref{TabXmaxShifts} for older and newer versions of the models, without taking statistical and systematic uncertainties into account.
Note the N and Fe nuclei represent groups for nuclei of similar masses.
Under the assumption of a constant \DeltaXmax modification in models, we see a general trend of suppression of protons and helium nuclei beyond the ankle energy ($\approx5$\,EeV).
An increase of the nitrogen fraction between the ankle and instep ($\approx13$\,EeV) energy \cite{PRL2020spectrum} is also noticeable for all modified model predictions.
In case of the iron nuclei, an increase of the relative fraction towards the highest energies is common for all predictions of modified models.

\begin{figure}
    \includegraphics[width=0.48\textwidth]{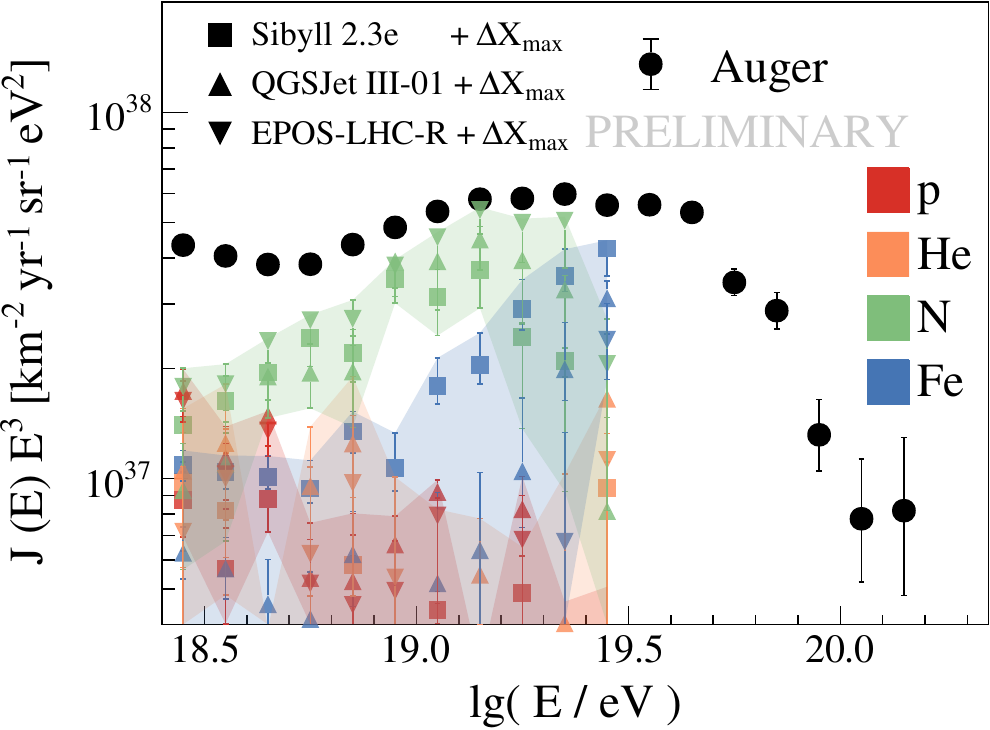}
    \hfill
    \includegraphics[width=0.48\textwidth]{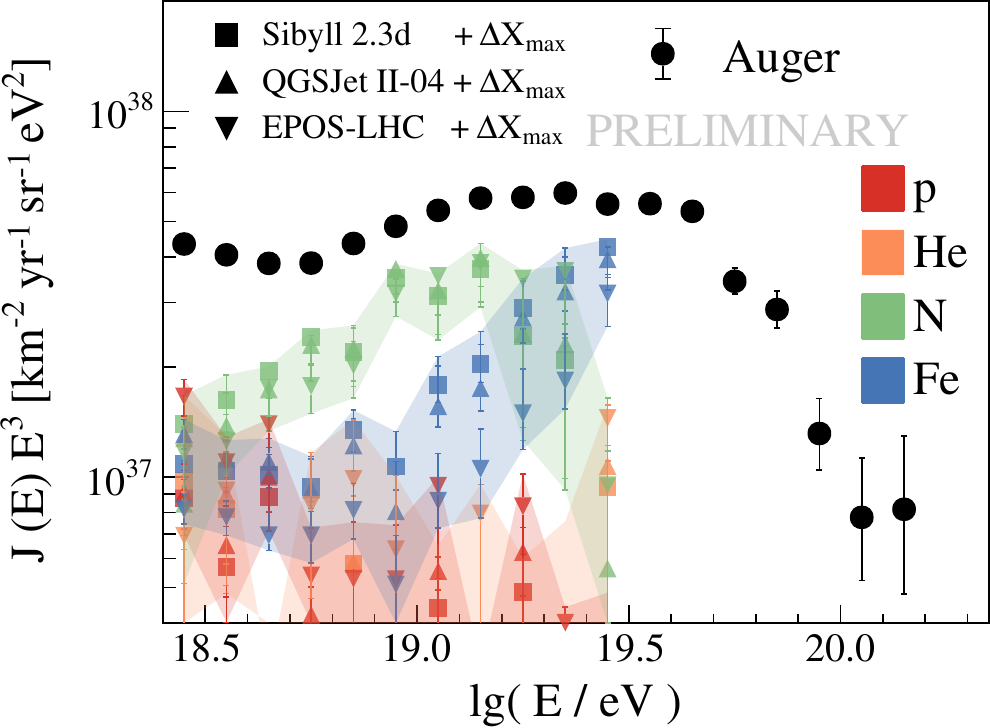}
    \caption{The total energy spectrum measured by the Pierre Auger Observatory \cite{SDEnergySpectrum2020} decomposed into four primary components using relative primary fractions shown in Fig.~\ref{fig:PrimFractionsVsEnergy} for older (right panel) and new (left panel) versions of the hadronic interaction models.}
    \label{fig:EnergySpectra}
\end{figure}

In Fig.~\ref{fig:EnergySpectra}, we multiply the total energy spectrum from \cite{SDEnergySpectrum2020} by the primary fractions obtained in Fig.~\ref{fig:PrimFractionsVsEnergy}.
It illustrates that despite a global shift towards deeper predictions on \Xmax with the new models, the remaining differences in other model predictions like \Xmax fluctuations or p-Fe difference in \avg{\Xmax} even increased. 
As a consequence, there is no convergence in the interpreted mass composition and thus of the individual energy spectra.
However, the connection between the instep feature in the energy spectrum and the start of fading of nitrogen nuclei from the beam, as proposed in \cite{HeavyMetal2025}, seems to be a common feature of all the model predictions, if the predicted \Xmax scale is shifted by \DeltaXmax. 
Note that this interpretation of instep as a transition between the dominance of different mass groups is consistent with what was found in \cite{CombinedFit2023}, where the instep feature was attributed to the change in dominance of He to N nuclei due to injection and propagation effect.

\begin{figure}
    \centering
    \includegraphics[width=0.6\textwidth]{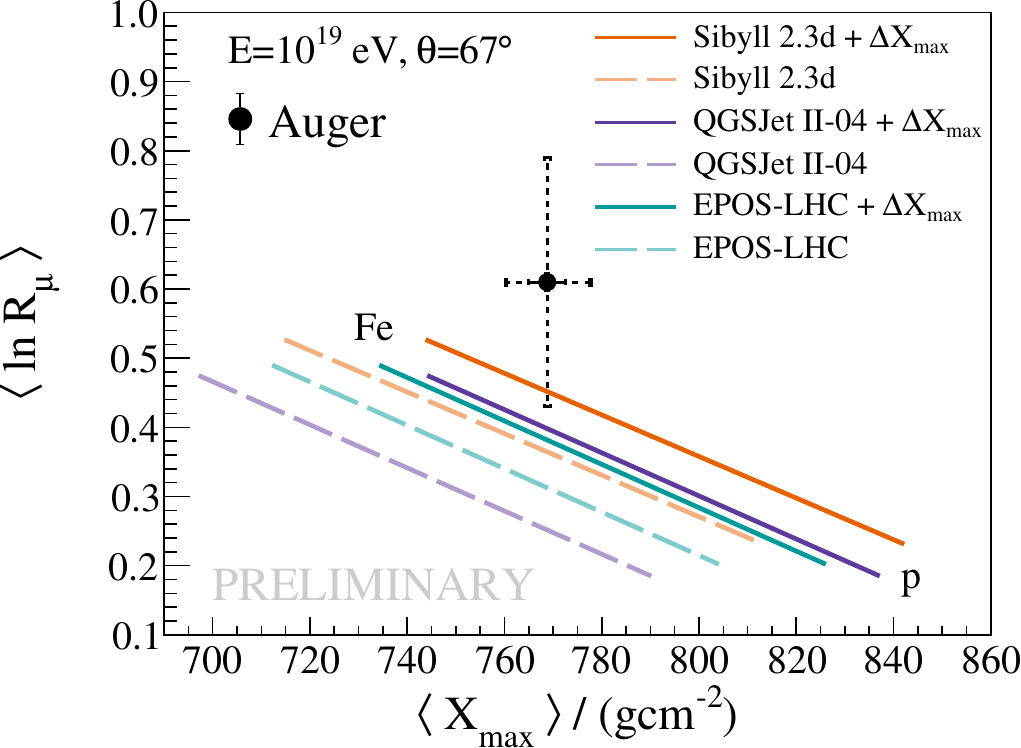}
    \caption{The muon scale ($R_{\mu}$) vs. \Xmax measurement at the Pierre Auger Observatory using inclined showers at energy $\sim$10\,EeV from \cite{MuonFluct2020}. We estimate predictions between p and Fe nuclei for original models of hadronic interactions (dashed lines) and for their shifted \Xmax predictions by \DeltaXmax (full lines).}
    \label{fig:MuonScale}
\end{figure}

Finally, we illustrate the alleviation of the muon problem in Fig.~\ref{fig:MuonScale} for the measurement in \cite{MuonFluct2020} for zenith angles $62^{\circ}\leq\theta\leq80^{\circ}$.
The original model predictions (dashed lines) are also shifted for \DeltaXmax obtained in \cite{PierreAuger:2024neu}.
The underestimation of the muon scale in the models is then reduced to about 15-25\%, which is in line with the values obtained in \cite{PierreAuger:2024neu} for the zenith angles $\theta\leq60^{\circ}$.
The new model \eposlhcr predicts more muons at larger zenith angles than its previous version, therefore better compatibility with measurement of the muon size in inclined showers is expected.

  \begin{table*}
    \centering
    \renewcommand{\arraystretch}{1.2}
      \caption{The \DeltaXmax values with statistical and systematical uncertainties found for older versions of hadronic interaction models in \cite{PierreAuger:2024neu} and here for the new models in the energy range $10^{18.5-19.0}$\,eV.}
      \begin{tabular}{l|ccc}
        & \eposlhc & \qgsii & \sib{2.3d} \\ 
        \DeltaXmax\, / (\gcm) & $22\pm3~^{+11}_{-14}$ & $47^{+2}_{-1}~^{+9}_{-11}$ & $29\pm2~^{+10}_{-13}$\\
        \hline\hline
        & \eposlhcr & \qgsiii & \sib{2.3e} \\ 
        \DeltaXmax\, / (\gcm) & $-1\pm2~^{+12}_{-16}$ & $18\pm1~^{+10}_{-13}$ & $28\pm2~^{+9}_{-14}$\\
      \end{tabular}
      \label{TabXmaxShifts}
  \end{table*}

\section{Summary}
The powerful combination of Surface and Fluorescence Detectors of the Pierre Auger Observatory allowed to test new versions of three models of hadronic interactions \eposlhcr, \qgsiii and \sib{2.3e} using the method from \cite{PierreAuger:2024neu}.
Although an improvement in the description of the measured \Xmax scale has been observed in the new versions of \epos and \qgs, all models are still unable to describe the measured data satisfactorily well in the energy range $10^{18.5-19.0}$\,eV.
All models seem to predict too hard spectra of muons causing less steep attenuation of the hadronic signal than is favored by the data.
Yet, interestingly, the primary fractions found to best describe the measurements are consistent between the older and new versions of the models, when \DeltaXmax and \RhadTheta modifications of the simulated templates are assumed.

The results of our studies in various energy ranges are compatible with no energy dependence of the modification parameters in the energy range $10^{18.4-19.5}$\,eV, which brings us to probe some basic phenomenology consequences regarding the energy spectrum and the lack of predicted muon scale compared to the direct measurement. 
For that, we assume a constant \DeltaXmax offset obtained in $10^{18.5-19.0}$\,eV and apply it in the model predictions on \Xmax to the full energy range $10^{18.4-19.5}$\,eV.
As a consequence of this assumption, for all models the protons and helium nuclei seem to be suppressed above the ankle energy.
The nitrogen nuclei increase their fraction in the primary beam above this energy up to the instep feature and start to steeply fade just beyond this energy.
Iron nuclei seem to increase their abundance towards the highest energies.
In case of the \DeltaXmax modifications, the problem of models from direct muon measurements at 10\,EeV is alleviated to the level of 15-25\% for older versions of the models, consistent with the result of \cite{PierreAuger:2024neu} at lower energy and zenith angles.




\small

\bibliographystyle{JHEP}
\bibliography{bibtex.bib}

%
%
%

\newpage

\par\noindent
\textbf{The Pierre Auger Collaboration}\\

\begin{wrapfigure}[8]{l}{0.12\linewidth}
\vspace{-2.9ex}
\includegraphics[width=0.98\linewidth]{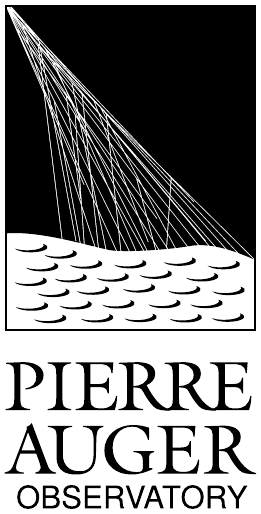}
\end{wrapfigure}
\begin{sloppypar}\noindent
\input{latex_authorlist_authors}
\end{sloppypar}

\input{latex_authorlist_institutions}
\input{acknowledgments}

\end{document}

%% file: latex_authorlist_authors.tex
A.~Abdul Halim$^{13}$,
P.~Abreu$^{70}$,
M.~Aglietta$^{53,51}$,
I.~Allekotte$^{1}$,
K.~Almeida Cheminant$^{78,77}$,
A.~Almela$^{7,12}$,
R.~Aloisio$^{44,45}$,
J.~Alvarez-Mu\~niz$^{76}$,
A.~Ambrosone$^{44}$,
J.~Ammerman Yebra$^{76}$
% ...

%% file: latex_authorlist_institutions.tex
\begin{description}[labelsep=0.2em,align=right,labelwidth=0.7em,labelindent=0em,leftmargin=2em,noitemsep,before={\renewcommand\makelabel[1]{##1 }}]
\item[$^{1}$] Centro At\'omico Bariloche and Instituto Balseiro (CNEA-UNCuyo-CONICET), San Carlos de Bariloche, Argentina
\item[$^{2}$] Departamento de F\'\i{}sica and Departamento de Ciencias de la Atm\'osfera y los Oc\'eanos, FCEyN, Universidad de Buenos Aires and CONICET, Buenos Aires, Argentina
\item[$^{3}$] IFLP, Universidad Nacional de La Plata and CONICET, La Plata, Argentina
\item[$^{4}$] Instituto de Astronom\'\i{}a y F\'\i{}sica del Espacio (IAFE, CONICET-UBA), Buenos Aires, Argentina
\item[$^{5}$] Instituto de F\'\i{}sica de Rosario (IFIR) -- CONICET/U.N.R.\ and Facultad de Ciencias Bioqu\'\i{}micas y Farmac\'euticas U.N.R., Rosario, Argentina
\item[$^{6}$] Instituto de Tecnolog\'\i{}as en Detecci\'on y Astropart\'\i{}culas (CNEA, CONICET, UNSAM), and Universidad Tecnol\'ogica Nacional -- Facultad Regional Mendoza (CONICET/CNEA), Mendoza, Argentina
\item[$^{7}$] Instituto de Tecnolog\'\i{}as en Detecci\'on y Astropart\'\i{}culas (CNEA, CONICET, UNSAM), Buenos Aires, Argentina
\item[$^{8}$] International Center of Advanced Studies and Instituto de Ciencias F\'\i{}sicas, ECyT-UNSAM and CONICET, Campus Miguelete -- San Mart\'\i{}n, Buenos Aires, Argentina
\item[$^{9}$] Laboratorio Atm\'osfera -- Departamento de Investigaciones en L\'aseres y sus Aplicaciones -- UNIDEF (CITEDEF-CONICET), Argentina
\item[$^{10}$] Observatorio Pierre Auger, Malarg\"ue, Argentina
\item[$^{11}$] Observatorio Pierre Auger and Comisi\'on Nacional de Energ\'\i{}a At\'omica, Malarg\"ue, Argentina
\item[$^{12}$] Universidad Tecnol\'ogica Nacional -- Facultad Regional Buenos Aires, Buenos Aires, Argentina
\item[$^{13}$] University of Adelaide, Adelaide, S.A., Australia
\item[$^{14}$] Universit\'e Libre de Bruxelles (ULB), Brussels, Belgium
\item[$^{15}$] Vrije Universiteit Brussels, Brussels, Belgium
\item[$^{16}$] Centro Brasileiro de Pesquisas Fisicas, Rio de Janeiro, RJ, Brazil
\item[$^{17}$] Centro Federal de Educa\c{c}\~ao Tecnol\'ogica Celso Suckow da Fonseca, Petropolis, Brazil
\item[$^{18}$] Instituto Federal de Educa\c{c}\~ao, Ci\^encia e Tecnologia do Rio de Janeiro (IFRJ), Brazil
\item[$^{19}$] Universidade de S\~ao Paulo, Escola de Engenharia de Lorena, Lorena, SP, Brazil
\item[$^{20}$] Universidade de S\~ao Paulo, Instituto de F\'\i{}sica de S\~ao Carlos, S\~ao Carlos, SP, Brazil
\item[$^{21}$] Universidade de S\~ao Paulo, Instituto de F\'\i{}sica, S\~ao Paulo, SP, Brazil
\item[$^{22}$] Universidade Estadual de Campinas (UNICAMP), IFGW, Campinas, SP, Brazil
\item[$^{23}$] Universidade Estadual de Feira de Santana, Feira de Santana, Brazil
\item[$^{24}$] Universidade Federal de Campina Grande, Centro de Ciencias e Tecnologia, Campina Grande, Brazil
\item[$^{25}$] Universidade Federal do ABC, Santo Andr\'e, SP, Brazil
\item[$^{26}$] Universidade Federal do Paran\'a, Setor Palotina, Palotina, Brazil
\item[$^{27}$] Universidade Federal do Rio de Janeiro, Instituto de F\'\i{}sica, Rio de Janeiro, RJ, Brazil
\item[$^{28}$] Universidad de Medell\'\i{}n, Medell\'\i{}n, Colombia
\item[$^{29}$] Universidad Industrial de Santander, Bucaramanga, Colombia
\item[$^{30}$] Charles University, Faculty of Mathematics and Physics, Institute of Particle and Nuclear Physics, Prague, Czech Republic
\item[$^{31}$] Institute of Physics of the Czech Academy of Sciences, Prague, Czech Republic
\item[$^{32}$] Palacky University, Olomouc, Czech Republic
\item[$^{33}$] CNRS/IN2P3, IJCLab, Universit\'e Paris-Saclay, Orsay, France
\item[$^{34}$] Laboratoire de Physique Nucl\'eaire et de Hautes Energies (LPNHE), Sorbonne Universit\'e, Universit\'e de Paris, CNRS-IN2P3, Paris, France
\item[$^{35}$] Univ.\ Grenoble Alpes, CNRS, Grenoble Institute of Engineering Univ.\ Grenoble Alpes, LPSC-IN2P3, 38000 Grenoble, France
\item[$^{36}$] Universit\'e Paris-Saclay, CNRS/IN2P3, IJCLab, Orsay, France
\item[$^{37}$] Bergische Universit\"at Wuppertal, Department of Physics, Wuppertal, Germany
\item[$^{38}$] Karlsruhe Institute of Technology (KIT), Institute for Experimental Particle Physics, Karlsruhe, Germany
\item[$^{39}$] Karlsruhe Institute of Technology (KIT), Institut f\"ur Prozessdatenverarbeitung und Elektronik, Karlsruhe, Germany
\item[$^{40}$] Karlsruhe Institute of Technology (KIT), Institute for Astroparticle Physics, Karlsruhe, Germany
\item[$^{41}$] RWTH Aachen University, III.\ Physikalisches Institut A, Aachen, Germany
\item[$^{42}$] Universit\"at Hamburg, II.\ Institut f\"ur Theoretische Physik, Hamburg, Germany
\item[$^{43}$] Universit\"at Siegen, Department Physik -- Experimentelle Teilchenphysik, Siegen, Germany
\item[$^{44}$] Gran Sasso Science Institute, L'Aquila, Italy
\item[$^{45}$] INFN Laboratori Nazionali del Gran Sasso, Assergi (L'Aquila), Italy
\item[$^{46}$] INFN, Sezione di Catania, Catania, Italy
\item[$^{47}$] INFN, Sezione di Lecce, Lecce, Italy
\item[$^{48}$] INFN, Sezione di Milano, Milano, Italy
\item[$^{49}$] INFN, Sezione di Napoli, Napoli, Italy
\item[$^{50}$] INFN, Sezione di Roma ``Tor Vergata'', Roma, Italy
\item[$^{51}$] INFN, Sezione di Torino, Torino, Italy
\item[$^{52}$] Istituto di Astrofisica Spaziale e Fisica Cosmica di Palermo (INAF), Palermo, Italy
\item[$^{53}$] Osservatorio Astrofisico di Torino (INAF), Torino, Italy
\item[$^{54}$] Politecnico di Milano, Dipartimento di Scienze e Tecnologie Aerospaziali , Milano, Italy
\item[$^{55}$] Universit\`a del Salento, Dipartimento di Matematica e Fisica ``E.\ De Giorgi'', Lecce, Italy
\item[$^{56}$] Universit\`a dell'Aquila, Dipartimento di Scienze Fisiche e Chimiche, L'Aquila, Italy
\item[$^{57}$] Universit\`a di Catania, Dipartimento di Fisica e Astronomia ``Ettore Majorana``, Catania, Italy
\item[$^{58}$] Universit\`a di Milano, Dipartimento di Fisica, Milano, Italy
\item[$^{59}$] Universit\`a di Napoli ``Federico II'', Dipartimento di Fisica ``Ettore Pancini'', Napoli, Italy
\item[$^{60}$] Universit\`a di Palermo, Dipartimento di Fisica e Chimica ''E.\ Segr\`e'', Palermo, Italy
\item[$^{61}$] Universit\`a di Roma ``Tor Vergata'', Dipartimento di Fisica, Roma, Italy
\item[$^{62}$] Universit\`a Torino, Dipartimento di Fisica, Torino, Italy
\item[$^{63}$] Benem\'erita Universidad Aut\'onoma de Puebla, Puebla, M\'exico
\item[$^{64}$] Unidad Profesional Interdisciplinaria en Ingenier\'\i{}a y Tecnolog\'\i{}as Avanzadas del Instituto Polit\'ecnico Nacional (UPIITA-IPN), M\'exico, D.F., M\'exico
\item[$^{65}$] Universidad Aut\'onoma de Chiapas, Tuxtla Guti\'errez, Chiapas, M\'exico
\item[$^{66}$] Universidad Michoacana de San Nicol\'as de Hidalgo, Morelia, Michoac\'an, M\'exico
\item[$^{67}$] Universidad Nacional Aut\'onoma de M\'exico, M\'exico, D.F., M\'exico
\item[$^{68}$] Institute of Nuclear Physics PAN, Krakow, Poland
\item[$^{69}$] University of \L{}\'od\'z, Faculty of High-Energy Astrophysics,\L{}\'od\'z, Poland
\item[$^{70}$] Laborat\'orio de Instrumenta\c{c}\~ao e F\'\i{}sica Experimental de Part\'\i{}culas -- LIP and Instituto Superior T\'ecnico -- IST, Universidade de Lisboa -- UL, Lisboa, Portugal
\item[$^{71}$] ``Horia Hulubei'' National Institute for Physics and Nuclear Engineering, Bucharest-Magurele, Romania
\item[$^{72}$] Institute of Space Science, Bucharest-Magurele, Romania
\item[$^{73}$] Center for Astrophysics and Cosmology (CAC), University of Nova Gorica, Nova Gorica, Slovenia
\item[$^{74}$] Experimental Particle Physics Department, J.\ Stefan Institute, Ljubljana, Slovenia
\item[$^{75}$] Universidad de Granada and C.A.F.P.E., Granada, Spain
\item[$^{76}$] Instituto Galego de F\'\i{}sica de Altas Enerx\'\i{}as (IGFAE), Universidade de Santiago de Compostela, Santiago de Compostela, Spain
\item[$^{77}$] IMAPP, Radboud University Nijmegen, Nijmegen, The Netherlands
\item[$^{78}$] Nationaal Instituut voor Kernfysica en Hoge Energie Fysica (NIKHEF), Science Park, Amsterdam, The Netherlands
\item[$^{79}$] Stichting Astronomisch Onderzoek in Nederland (ASTRON), Dwingeloo, The Netherlands
\item[$^{80}$] Universiteit van Amsterdam, Faculty of Science, Amsterdam, The Netherlands
\item[$^{81}$] Case Western Reserve University, Cleveland, OH, USA
\item[$^{82}$] Colorado School of Mines, Golden, CO, USA
\item[$^{83}$] Department of Physics and Astronomy, Lehman College, City University of New York, Bronx, NY, USA
\item[$^{84}$] Michigan Technological University, Houghton, MI, USA
\item[$^{85}$] New York University, New York, NY, USA
\item[$^{86}$] University of Chicago, Enrico Fermi Institute, Chicago, IL, USA
\item[$^{87}$] University of Delaware, Department of Physics and Astronomy, Bartol Research Institute, Newark, DE, USA
\item[] -----
\item[$^{a}$] Max-Planck-Institut f\"ur Radioastronomie, Bonn, Germany
\item[$^{b}$] also at Kapteyn Institute, University of Groningen, Groningen, The Netherlands
\item[$^{c}$] School of Physics and Astronomy, University of Leeds, Leeds, United Kingdom
\item[$^{d}$] Fermi National Accelerator Laboratory, Fermilab, Batavia, IL, USA
\item[$^{e}$] Pennsylvania State University, University Park, PA, USA
\item[$^{f}$] Colorado State University, Fort Collins, CO, USA
\item[$^{g}$] Louisiana State University, Baton Rouge, LA, USA
\item[$^{h}$] now at Graduate School of Science, Osaka Metropolitan University, Osaka, Japan
\item[$^{i}$] Institut universitaire de France (IUF), France
\item[$^{j}$] now at Technische Universit\"at Dortmund and Ruhr-Universit\"at Bochum, Dortmund and Bochum, Germany
\end{description}

%% file: acknowledgments.tex
\section*{Acknowledgments}

\begin{sloppypar}
The successful installation, commissioning, and operation of the Pierre
Auger Observatory would not have been possible without the strong
commitment and effort from the technical and administrative staff in
Malarg\"ue. We are very grateful to the following agencies and
organizations for financial support:
\end{sloppypar}

\begin{sloppypar}
Argentina -- Comisi\'on Nacional de Energ\'\i{}a At\'omica; Agencia Nacional de
Promoci\'on Cient\'\i{}fica y Tecnol\'ogica (ANPCyT); Consejo Nacional de
Investigaciones Cient\'\i{}ficas y T\'ecnicas (CONICET); Gobierno de la
Provincia de Mendoza; Municipalidad de Malarg\"ue; NDM Holdings and Valle
Las Le\~nas; in gratitude for their continuing cooperation over land
access; Australia -- the Australian Research Council; Belgium -- Fonds
de la Recherche Scientifique (FNRS); Research Foundation Flanders (FWO),
Marie Curie Action of the European Union Grant No.~101107047; Brazil --
Conselho Nacional de Desenvolvimento Cient\'\i{}fico e Tecnol\'ogico (CNPq);
Financiadora de Estudos e Projetos (FINEP); Funda\c{c}\~ao de Amparo \`a
Pesquisa do Estado de Rio de Janeiro (FAPERJ); S\~ao Paulo Research
Foundation (FAPESP) Grants No.~2019/10151-2, No.~2010/07359-6 and
No.~1999/05404-3; Minist\'erio da Ci\^encia, Tecnologia, Inova\c{c}\~oes e
Comunica\c{c}\~oes (MCTIC); Czech Republic -- GACR 24-13049S, CAS LQ100102401,
MEYS LM2023032, CZ.02.1.01/0.0/0.0/16{\textunderscore}013/0001402,
CZ.02.1.01/0.0/0.0/18{\textunderscore}046/0016010 and
CZ.02.1.01/0.0/0.0/17{\textunderscore}049/0008422 and CZ.02.01.01/00/22{\textunderscore}008/0004632;
France -- Centre de Calcul IN2P3/CNRS; Centre National de la Recherche
Scientifique (CNRS); Conseil R\'egional Ile-de-France; D\'epartement
Physique Nucl\'eaire et Corpusculaire (PNC-IN2P3/CNRS); D\'epartement
Sciences de l'Univers (SDU-INSU/CNRS); Institut Lagrange de Paris (ILP)
Grant No.~LABEX ANR-10-LABX-63 within the Investissements d'Avenir
Programme Grant No.~ANR-11-IDEX-0004-02; Germany -- Bundesministerium
f\"ur Bildung und Forschung (BMBF); Deutsche Forschungsgemeinschaft (DFG);
Finanzministerium Baden-W\"urttemberg; Helmholtz Alliance for
Astroparticle Physics (HAP); Helmholtz-Gemeinschaft Deutscher
Forschungszentren (HGF); Ministerium f\"ur Kultur und Wissenschaft des
Landes Nordrhein-Westfalen; Ministerium f\"ur Wissenschaft, Forschung und
Kunst des Landes Baden-W\"urttemberg; Italy -- Istituto Nazionale di
Fisica Nucleare (INFN); Istituto Nazionale di Astrofisica (INAF);
Ministero dell'Universit\`a e della Ricerca (MUR); CETEMPS Center of
Excellence; Ministero degli Affari Esteri (MAE), ICSC Centro Nazionale
di Ricerca in High Performance Computing, Big Data and Quantum
Computing, funded by European Union NextGenerationEU, reference code
CN{\textunderscore}00000013; M\'exico -- Consejo Nacional de Ciencia y Tecnolog\'\i{}a
(CONACYT) No.~167733; Universidad Nacional Aut\'onoma de M\'exico (UNAM);
PAPIIT DGAPA-UNAM; The Netherlands -- Ministry of Education, Culture and
Science; Netherlands Organisation for Scientific Research (NWO); Dutch
national e-infrastructure with the support of SURF Cooperative; Poland
-- Ministry of Education and Science, grants No.~DIR/WK/2018/11 and
2022/WK/12; National Science Centre, grants No.~2016/22/M/ST9/00198,
2016/23/B/ST9/01635, 2020/39/B/ST9/01398, and 2022/45/B/ST9/02163;
Portugal -- Portuguese national funds and FEDER funds within Programa
Operacional Factores de Competitividade through Funda\c{c}\~ao para a Ci\^encia
e a Tecnologia (COMPETE); Romania -- Ministry of Research, Innovation
and Digitization, CNCS-UEFISCDI, contract no.~30N/2023 under Romanian
National Core Program LAPLAS VII, grant no.~PN 23 21 01 02 and project
number PN-III-P1-1.1-TE-2021-0924/TE57/2022, within PNCDI III; Slovenia
-- Slovenian Research Agency, grants P1-0031, P1-0385, I0-0033, N1-0111;
Spain -- Ministerio de Ciencia e Innovaci\'on/Agencia Estatal de
Investigaci\'on (PID2019-105544GB-I00, PID2022-140510NB-I00 and
RYC2019-027017-I), Xunta de Galicia (CIGUS Network of Research Centers,
Consolidaci\'on 2021 GRC GI-2033, ED431C-2021/22 and ED431F-2022/15),
Junta de Andaluc\'\i{}a (SOMM17/6104/UGR and P18-FR-4314), and the European
Union (Marie Sklodowska-Curie 101065027 and ERDF); USA -- Department of
Energy, Contracts No.~DE-AC02-07CH11359, No.~DE-FR02-04ER41300,
No.~DE-FG02-99ER41107 and No.~DE-SC0011689; National Science Foundation,
Grant No.~0450696, and NSF-2013199; The Grainger Foundation; Marie
Curie-IRSES/EPLANET; European Particle Physics Latin American Network;
and UNESCO.
\end{sloppypar}